\def\year{2020}\relax
\documentclass[letterpaper]{article} % DO NOT CHANGE THIS
\usepackage{aaai20}  % DO NOT CHANGE THIS
\usepackage{times}  % DO NOT CHANGE THIS
\usepackage{helvet} % DO NOT CHANGE THIS
\usepackage{courier}  % DO NOT CHANGE THIS
\usepackage[hyphens]{url}  % DO NOT CHANGE THIS
\usepackage{graphicx} % DO NOT CHANGE THIS
\usepackage{graphicx}  % DO NOT CHANGE THIS
\usepackage{multirow}
\usepackage{booktabs}
\usepackage{amsmath}

\title{Segmenting Medical MRI via Recurrent Decoding Cell}
\author{Ying Wen,\textsuperscript{\rm 1} Kai Xie,\textsuperscript{\rm 2} Lianghua He\textsuperscript{\rm 3}\thanks{Corresponding author. Email: helianghua@tongji.edu.cn.}\\
\textsuperscript{\rm 1}School of Communication and Electronic Engineering \&\\
 Shanghai Key Laboratory of Multidimensional Information Processing, East China Normal University, Shanghai, China\\
\textsuperscript{\rm 2}School of Computer Science and Technology, East China Normal University, Shanghai, China\\ 
\textsuperscript{\rm 3}Department of Computer Science and Technology, Tongji University, Shanghai, China\\ 
}

\begin{document}
 	
\maketitle

\begin{abstract}
	The encoder-decoder networks are commonly used in medical image segmentation due to their remarkable performance in hierarchical feature fusion. However, the expanding path for feature decoding and spatial recovery does not consider the long-term dependency when fusing feature maps from different layers, and the universal encoder-decoder network does not make full use of the multi-modality information to improve the network robustness especially for segmenting medical MRI. In this paper, we propose a novel feature fusion unit called Recurrent Decoding Cell (RDC) which leverages convolutional RNNs to memorize the long-term context information from the previous layers in the decoding phase. An encoder-decoder network, named Convolutional Recurrent Decoding Network (CRDN), is also proposed based on RDC for segmenting multi-modality medical MRI. CRDN adopts CNN backbone to encode image features and decode them hierarchically through a chain of RDCs to obtain the final high-resolution score map. The evaluation experiments on BrainWeb, MRBrainS and HVSMR datasets demonstrate that the introduction of RDC effectively improves the segmentation accuracy as well as reduces the model size, and the proposed CRDN owns its robustness to image noise and intensity non-uniformity in medical MRI. 
\end{abstract}

\section{Introduction}
\label{sec:intro}
	Magnetic Resonance Imaging (MRI) plays a pivotal role in the analysis of neuroscience and the diagnosis of disease. The accurate segmentation of medical MRI enables doctors and researchers to obtain the anatomical information about different parts of biological tissues. However, the traditional way of pixel-level annotations by experts is tedious and time-consuming, thus methods for automatic MRI segmentation gain interest. Clustering-based methods \cite{dunn1973fuzzy,gong2012fuzzy} have shown satisfying segmentation for certain entire and high-contrast images like MR brain slices. In spite of this, these methods consume much time for iterations and are not robust enough for inhomogeneous intensity and image noise in MRI. These years deep learning based methods have shown its superiority in feature extraction, among which fully convolutional network (FCN) \cite{long2015fully} is first proposed for the task of semantic segmentation. Since then, FCN-like networks \cite{ronneberger2015u,dolz2018hyperdense,sinha2019multi} have successfully been applied to medical image segmentation due to their remarkable segmentation accuracy as well as the stability and robustness to inhomogeneous intensity.
	
	There still exist three main challenges for medical image segmentation. Firstly, the importance of hierarchical feature fusion. For medical images, the semantic information extracted from the deep layers is relatively simpler  the spatial information extracted from the shallow layers is more helpful compared with natural image segmentation. The biological details and spatial information for labeling the region of interest accurately count a lot, which in turn requires the designed network to own a better decoding capability for hierarchical feature fusion and spatial recovery. Secondly, the use of multi-modality information. Medical images, especially MRI images, often have multi-modality scans (such as T1, T2 and PD) obtained from different devices, and different modalities respond differently to various tissues. Hence, leveraging multi-modality information is beneficial to deal with the insufficient tissue contrast problem and improve segmentation accuracy \cite{tseng2017joint}. Thirdly, the robustness of networks. Sufficient training samples are not easy to obtain for medical image segmentation, thus the trained model may easily experience overfitting and be sensitive to image noise and intensity non-uniformity fields, and this requires the robustness of network design.
	
	For hierarchical feature fusion, the encoder-decoder structure has exhibited its superiority and is widely  medical image segmentation. Models like U-Net \cite{ronneberger2015u} and its variants \cite{milletari2016v,zhou2018unetpp} encode information from different resolution of feature maps. Feature maps from deeper layers of a CNN backbone encode higher-level semantic information and context contained in the large receptive field, and the shallow layers encode biological appearance and spatial information in a relatively small receptive field. The decoders of these networks utilize the encoded information from all layers, and combine the lower-level features and higher-level features step by step to gradually recover the input spatial resolution. However, many decoders only use concatenation or element-wise summation for the fusion of feature information across layers. This may neglect the long-term memory of the former layers, which is to say, although feature maps with higher resolution are utilized in each decoding stage, the last fused feature map for prediction could still lose the information from the early fusion stage since the operations for hierarchical feature fusion are not capable enough in memory. 
	
	Inspired by the above analysis of segmenting medical MRI, in this paper, we propose a Recurrent Decoding Cell (RDC) for better hierarchical feature fusion with its strong ability to memorize long-term context information through the decoding pathway. The RDC is a parameter-sharing unit in each fusion stage which combines the current score map of low resolution with the squeezed feature map of high resolution. The convolutional RNN is introduced in each RDC unit for long-term spatial and semantic information fusion. Three types of RDCs are implemented in our experiment according to RNN and its variants, namely RDC with basic convolutional RNN (ConvRNN), RDC with ConvLSTM, and RDC with ConvGRU. Moreover, for multi-modality training and robustness to intensity-related artifacts, we also propose a Convolutional Recurrent Decoding Network (CRDN) based on RDC for segmenting multi-modality medical MRI. The CRDN can receive multi-modality images as input, and encodes the semantic and spatial information through a CNN backbone to generate hierarchical feature maps, then the RDC-based decoder improves an initialized score map through the long-term memory path to generate hierarchical score maps. The final score map with the same resolution as the input image is considered as the final predication. We conduct experiments on two brain segmentation datasets and one cardiovascular MRI dataset, the BrainWeb \cite{cocosco1997brainweb}, MRBrainS \cite{mendrik2015mrbrains} and HVSMR \cite{pace2015interactive}. Several experimental results reveal that our CRDN enjoys segmentation accuracy gains compared with other excellent encoder-decoder networks, and our model also owns its robustness to image noise and intensity non-uniformity in MRI. Moreover, CRDN achieves smaller model size due to the shared parameters in RDC. Our contributions are as follows:
	
	\begin{itemize}
		\item We propose a new feature fusion unit called Recurrent Decoding Cell (RDC), which leverages the ability of convolutional RNN in memorizing long-term context information. The parameters in RDC are shared in each hierarchical stage, therefore, it is a flexible module and can be added into any encoder-decoder segmentation network to help reduce model size. 
		\item We propose a Convolutional Recurrent Decoding Network (CRDN) based on RDC for segmenting multi-modality medical MRI. CRDN utilizes CNN backbone as the feature encoder and RDC-based decoder to form an end-to-end segmentation network. CRDN effectively increases the segmentation accuracy and shows its robustness in image noise and intensity non-uniformity.
	\end{itemize}

\begin{figure}[htb]
	\centering
	\includegraphics[width=1.0\linewidth]{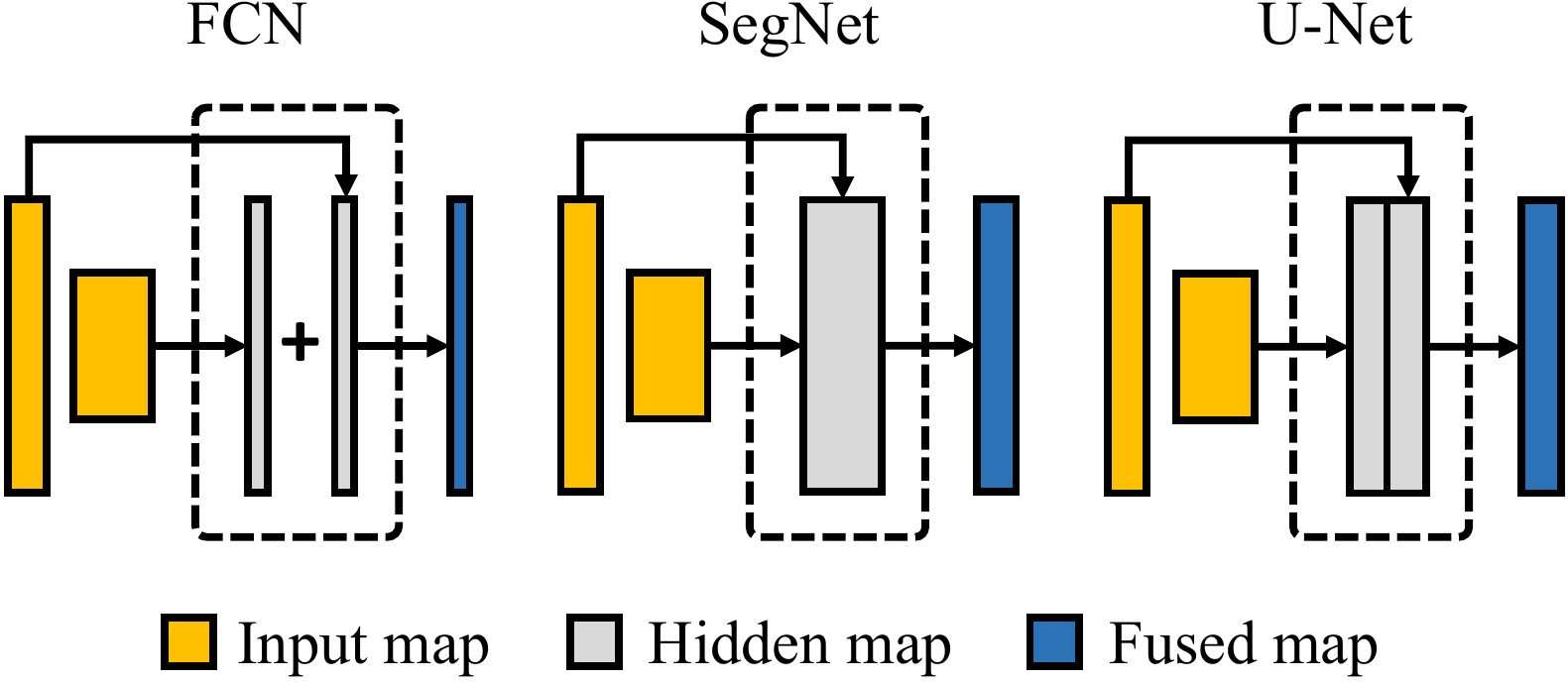}
	\caption{Abstract illustrations of the feature fusion unit in FCN, SegNet and U-Net. The two yellow boxes represent two multi-channel feature maps. The gray boxes within the dashed rectangle are hidden maps in the feature fusion unit. The blue boxes are fused maps. The solid lines with arrows correspond to different operations for feature map squeezing or upsampling.}
	\label{fig:threefuse}
\end{figure}

\begin{figure*}[htb]
	\centering
	\includegraphics[width=1.00\linewidth]{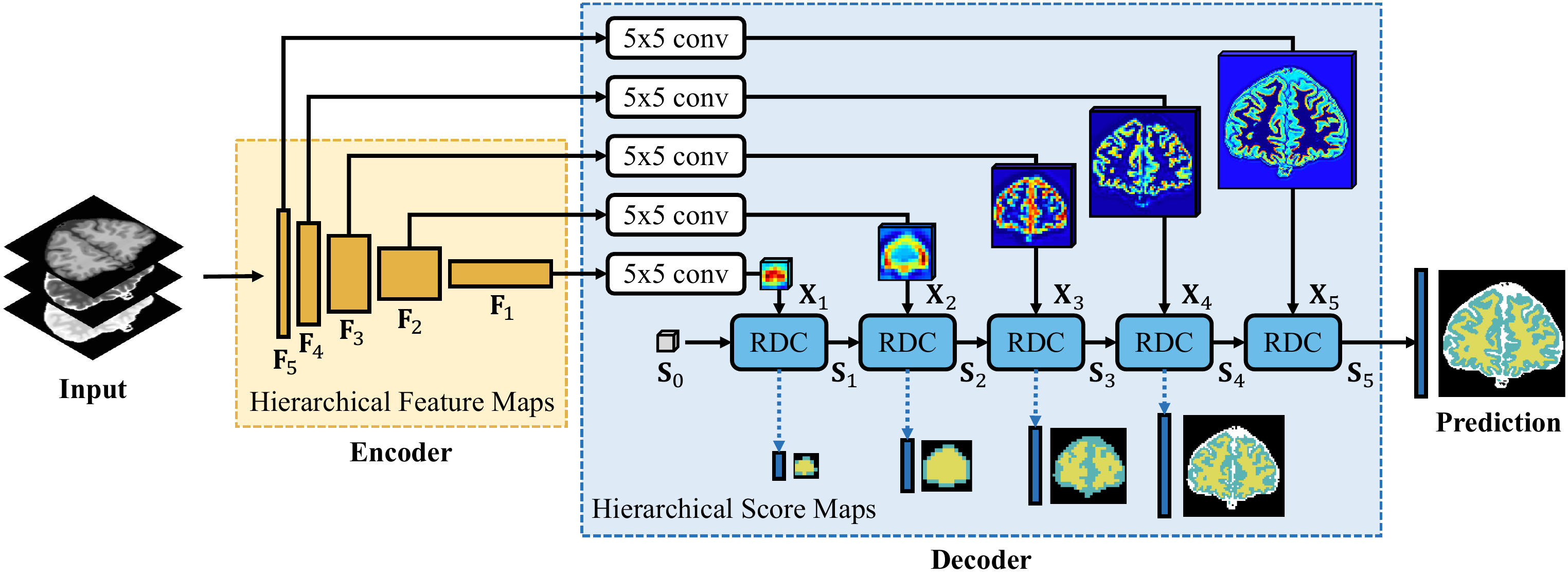}
	\caption{Illustration of the proposed Convolutional Recurrent Decoding Network.}
	\label{fig:network}
\end{figure*}

\section{Related Work} 
	\paragraph{Encoder-Decoder Structure for Medical Image Segmentation} 
	Hierarchical feature fusion is helpful for precious boundary adherence in medical image segmentation. The encoder-decoder structures fuse the two multi-channel feature maps with different spatial resolutions in each feature fusion unit. Figure \ref{fig:threefuse} shows an abstract illustration of three popular encoder-decoder networks in feature fusion. Feature maps from different layers in FCN \cite{long2015fully} are first squeezed by convolution to produce score maps of different resolutions, the score map with lower resolution is upsampled and added to the score map with higher resolution to form the fused map. SegNet \cite{badrinarayanan2017segnet} adopts unpooling and convolution to expand the feature map according to the maxpooling indices obtained from the higher resolution map. U-Net \cite{ronneberger2015u} adopts transposed convolution to squeeze the lower resolution feature map as well as expands their spatial size the same as the higher resolution feature map, and the two maps are then concatenated to form the fused map. Many other encoder-decoder methods are also proposed based on the above three designs for segmenting medical images. InvertedNet \cite{novikov2018fully} is improved by U-Net, and it utilizes delayed subsampling to learn higher resolution features and has fewer parameters to prevent overfitting. CE-Net \cite{gu2019net} adds a context extractor block between the encoder and the decoder to reduce the information loss caused by pooling and convolution.

	\paragraph{Convolutional Recurrent Neural Networks}
	The recurrent neural networks, especially LSTM \cite{hochreiter1997long} and GRU \cite{cho2014learning}, have natural advantages in memorizing long-term context information. The convolutional version of RNN further extends this ability to 2D image sequence. Shi et al. \cite{xingjian2015convolutional,shi2017deep} first applied the convolution-based RNN to precipitation nowcasting, which proves the powerful ability of ConvLSTM and ConvGRU for capturing spatiotemporal correlations. Bo et al. \cite{pang2019deep} designed a representation bridge module based on convolutional RNNs for visual sequential applications, and achieves state-of-the-art performance in most visual sequential tasks. However, convolutional RNN has not been applied to feature fusion in medical image segmentation. Hence, we would like to make full use of its advantage for the fusion of long-term spatial information between the feature maps of different layers.

\section{Method}
\label{sec:method}
	We propose a medical MRI segmentation network called Convolutional Recurrent Decoding Network (CRDN). In this network, a novel feature fusion unit called Recurrent Decoding Cell (RDC) is also proposed. CRDN is an encoder-decoder network which receives multi-modality images as input and generates segmentation inference through a CNN backbone encoder and the RDC-based decoder. The RDC is a flexible and parameter-sharing unit used in CRDN for hierarchical feature fusion, in which a convolutional RNN is used to combine the spatial and semantic information between feature maps. The final segmentation result is obtained from the last fused score map decoded by RDC. In this section, we introduce our CRDN and the RDC unit in more detail.

	\subsection{Convolutional Recurrent Decoding Network}
	\label{subsec:crdn}
		The proposed CRDN is an end-to-end segmentation pipeline which takes multi-modality images as input and produces per-pixel segmentation inference for each tissue. CRDN consists of two phases: the CNN backbone is utilized as the encoder to extract feature maps for hierarchical feature learning and the proposed recurrent decoding cell (RDC) is designed as the decoder to gradually recover the spatial resolution, and its overall pipeline is shown in Figure \ref{fig:network}.  
		
		Given a multi-modality medical image $\mathbf{I}$, a collection of hierarchical feature maps $\{\mathbf{F}_i\}_{i=1}^L$ with different resolutions is initially produced by a CNN backbone like VGG or ResNet, where $L$ is the number of layers of CNN hierarchy. $\mathbf{F}_i$ encodes the multi-scale context information in a coarse-to-fine manner, and the resolution of each feature map halves while the number of channels increase through the CNN encoding flow. Here we remark $\mathbf{F}_1$ has the lowest resolution. Next, feature maps $\{\mathbf{F}_1,…,\mathbf{F}_L\}$ are further squeezed into $C$ channels, where $C$ is the number of segmentation classes. It is done through a distinct $5 \times 5$ convolution filter with zero padding equals $2$, following by a ReLU activation, which is written as
		
		\begin{eqnarray}
		\mathbf{X}_i=ReLU(\mathbf{F}_i \otimes \psi_i)
		\end{eqnarray}
		where $\mathbf{X}_i$ is a $C$-dimensional feature map, $\psi_i$ is the convolution kernel parameters in the $i$th layer. The reduction of channels produces per-class feature maps from different scales and effectively reduces model size in the decoding phase. 
		
		The decoder consists of a $L$-stage recurrent decoding chain. It gradually incorporates feature maps of different scales with score maps to decode the final prediction. Specifically, starting from the initial score map $\mathbf{S}_0$, $L$ RDC units are followed to recover the final prediction score map. The previous score map $\mathbf{S}_{i-1}$ with low spatial resolution and the current feature map $\mathbf{X}_i$ with relatively high spatial resolution are fed into the current RDC, yielding the current score map $\mathbf{S}_i$ with the same resolution as $\mathbf{X}_i$. This can be written as follows
		
		\begin{eqnarray}
		\mathbf{S}_i=RDC(\mathbf{S}_{i-1},\mathbf{X}_i;\phi)
		\end{eqnarray}
		where RDC is the proposed unit for hierarchical feature refinement, $\phi$ is the shared parameters, $\mathbf{S}_0$ is initialized as a $C$-dimensional zero tensor as the initial score map. Along the RDC chain, the decoding flow learns to assimilate and memorize features of different scales and produces hierarchical score maps $\{\mathbf{S}_1,…,\mathbf{S}_L\}$. Among them, the current score map is twice the size as the previous one, and contains richer spatial information as well as maintaining semantic information.
		
		Finally, $\mathbf{S}_L$ is treated as the final score map. The loss function for a single map prediction is defined as the sum of cross-entropy losses at individual pixels between the ground truth and $\mathbf{S}_L$ through epochs of back-propagation.

	\subsection{Recurrent Decoding Cell}
	\label{subsec:rdc}
		The Recurrent Decoding Cell is a feature fusion unit that can memorize the long-term context information to refine the current score map. The intuition here is that the collection of score maps can be treated as a coarse-to-fine sequence, the adjacent score maps have temporal and spatial correlations to each other, and the information helpful for the final segmentation is propagated through a chain of RDCs.
		
\begin{figure}[htb]
	\centering
	\includegraphics[width=0.9\linewidth]{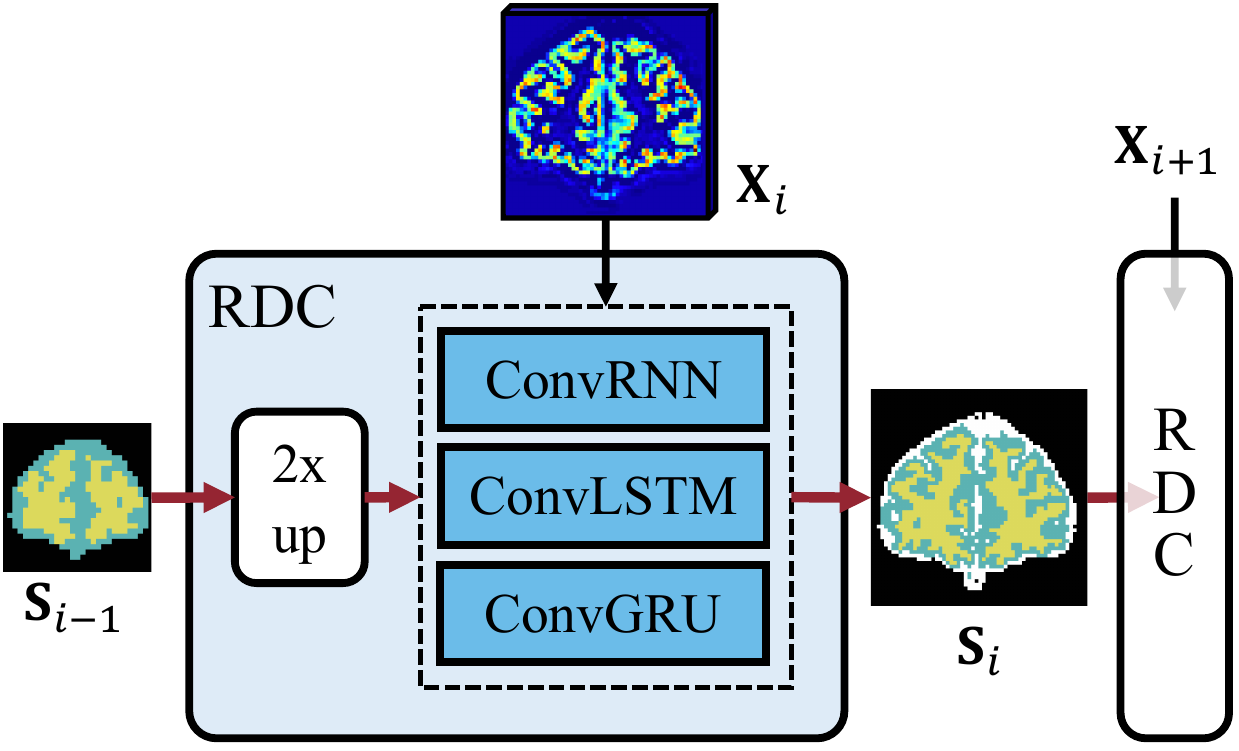}
	\caption{The structure of RDC. The previous score map, the current input feature map, the current score map and the next input feature map are denoted as $\mathbf{S}_{i-1}$, $\mathbf{X}_i$, $\mathbf{S}_i$, $\mathbf{X}_{i+1}$, respectively. One of the convolutional RNNs in the dashed box is used for feature fusion.}
	\label{fig:rdc}
\end{figure}

		Figure \ref{fig:rdc} illustrates the structure of RDC. In each RDC unit, the previous score map $\mathbf{S}_{i-1}$, which can be treated as the hidden state of an RNN cell, is refined with the current input $\mathbf{X}_i$, generating the current new score map $\mathbf{S}_i$ as the input of the following RDC. Specifically, we first upsample the score map $\mathbf{S}_{i-1}$ to the same spatial dimension as $\mathbf{X}_i$, and this can be done through either bilinear interpolation or learnable transposed convolution. Then, the upsampled score map and the current feature input are fed into a convolutional RNN cell for feature decoding. According to different types of RNNs, three types of RDCs are defined as follows.

		\paragraph{ConvRNN Decoding.} Here we denote ‘ConvRNN’ as the basic convolutional RNN utilized in our RDC unit, which can be formulated as
		
		\begin{eqnarray}
		\mathbf{S}_i=\sigma(\textbf{\emph{W}}_{s} \otimes T(\mathbf{S}_{i-1})+\textbf{\emph{W}}_{x} \otimes \mathbf{X}_{i})
		\end{eqnarray}
		where $\textbf{\emph{W}}$ are the weight matrices learned from the network and the bias terms are omitted for notational simplicity. $\otimes$ is the convolution operation. $T(.)$ denotes the above mentioned upsampling operation. $\sigma(.)$ is an activation function, and we use ReLU in practice. From another perspective, we can also consider the ConvRNN cell as a concatenate-conv-ReLU operation used in U-Net, and it is a simple but efficient way of feature fusion. We denote this unit as ‘RDC-ConvRNN’ in this paper.

		\paragraph{ConvLSTM Decoding.} A ConvLSTM cell computes four different gates $\mathbf{G}^{i}$, $\mathbf{G}^{f}$, $\mathbf{G}^{o}$, $\mathbf{G}^{g}$ to specifically decide whether and how much to propagate both semantic and spatial information to the next RDC unit. This can be formulated as
		
		\begin{equation}\label{eq:convlstm}
		\begin{split}
		\mathbf{G}^{i} &= \sigma(\textbf{\emph{W}}_{xi} \otimes \mathbf{X}_i + \textbf{\emph{W}}_{si} \otimes T(\mathbf{S}_{i-1}))\\
		\mathbf{G}^{f} &= \sigma(\textbf{\emph{W}}_{xf} \otimes \mathbf{X}_i + \textbf{\emph{W}}_{sf} \otimes T(\mathbf{S}_{i-1}))\\
		\mathbf{G}^{o} &= \sigma(\textbf{\emph{W}}_{xo} \otimes \mathbf{X}_i + \textbf{\emph{W}}_{so} \otimes T(\mathbf{S}_{i-1}))\\
		\mathbf{G}^{g} &= \delta(\textbf{\emph{W}}_{xg} \otimes \mathbf{X}_i + \textbf{\emph{W}}_{sg} \otimes T(\mathbf{S}_{i-1}))\\
		\mathbf{C}_{i} &= \mathbf{G}^f \odot T(\mathbf{C}_{i-1}) + \mathbf{G}^i \odot \mathbf{G}^g\\
		\mathbf{S}_{i} &= \mathbf{G}^o \odot \delta(\mathbf{C}_i)
		\end{split}
		\end{equation}
		where $\mathbf{C}_{i}$ indicate the cell state of ConvLSTM. $\textbf{\emph{W}}$ are learnable weight matrix. $\odot$ denotes the point-wise product. $T(.)$ is the upsampling operation. $\sigma(.)$ and $\delta(.)$ are two activation functions, and we use RuLU and Tanh respectively. Every time a new input arrives, the four gates $\mathbf{G}^{i}$, $\mathbf{G}^{f}$, $\mathbf{G}^{o}$, $\mathbf{G}^{g}$ control whether to write to the cell, whether to erase cell, how much to reveal cell and how much to write to the cell, respectively. Here we denote this unit as ‘RDC-ConvLSTM’.

		\paragraph{ConvGRU Decoding.} Similar to the ConvLSTM cell, ConvGRU computes two gates, namely reset gate and update gate, to decide whether to clear or update the visual information from the previous score map to the next RDC unit. This can be formulated as

		\begin{equation}\label{eq:convgru}
		\begin{split}
		\mathbf{G}^r &= \sigma(\textbf{\emph{W}}_{xr} \otimes \mathbf{X}_{i} + \textbf{\emph{W}}_{sr} \otimes T(\mathbf{S}_{i-1}))\\
		\mathbf{G}^z &= \sigma(\textbf{\emph{W}}_{xz} \otimes \mathbf{X}_{i} + \textbf{\emph{W}}_{sz} \otimes T(\mathbf{S}_{i-1}))\\
		\widetilde{\mathbf{S}}_i &= \delta(\textbf{\emph{W}}_{xs} \odot \mathbf{X}_{i} + \mathbf{G}^r \odot (\textbf{\emph{W}}_{ss} \otimes T(\mathbf{S}_{i-1})))\\
		\mathbf{S}_i &= \mathbf{G}^z \odot T(\mathbf{S}_{i-1}) + (1-\mathbf{G}^z) \odot \widetilde{\mathbf{S}}_i
		\end{split}
		\end{equation}
		where $\mathbf{G}^{r}$, $\mathbf{G}^{z}, \widetilde{\mathbf{S}}_i$ denote the reset gate, update gate and new information, respectively. $\textbf{\emph{W}}$ are learnable weight matrix. The reset gate controls whether to clear the previous state $\mathbf{S}_{i-1}$ and the update state controls how much the new information will be written to the output Score map $\mathbf{S}_i$. ConvGRU is relatively easy to train compared with ConvLSTM in practice \cite{ballas2015delving} and is effective to prevent vanishing or exploding of gradient. This decoding unit is denoted as ‘RDC-ConvGRU’.
		
		Along the RDC chain, since the number of channels of score maps from each stage remain the same, the RDC can share its parameters in the decoding phase, which makes it possible to use RDC recurrently and effectively controls the model size.

\section{Experiments}
\label{exp}

	In this section, we quantitatively evaluate the proposed CRDN for medical image segmentation. We test on two brain datasets and one cardiovascular MRI dataset: the BrainWeb dataset \cite{cocosco1997brainweb}, the MICCAI 2013 MRBrainS Challenge dataset \cite{mendrik2015mrbrains} and the HVSMR 2016 Challenge dataset \cite{pace2015interactive}. We first introduce the three datasets and the implementation details. Next, an ablation study is conducted to test the performance on different combinations of CNN backbones and RDCs. Then, we evaluate our CRDN in comparison with other encoder-decoder networks, i.e., FCN, SegNet and U-Net. Finally, we evaluate the robustness of CRDN when images are affected by noise and intensity non-uniformity.
	
\begin{figure}[htb]
	\centering
	\includegraphics[width=1.0\linewidth]{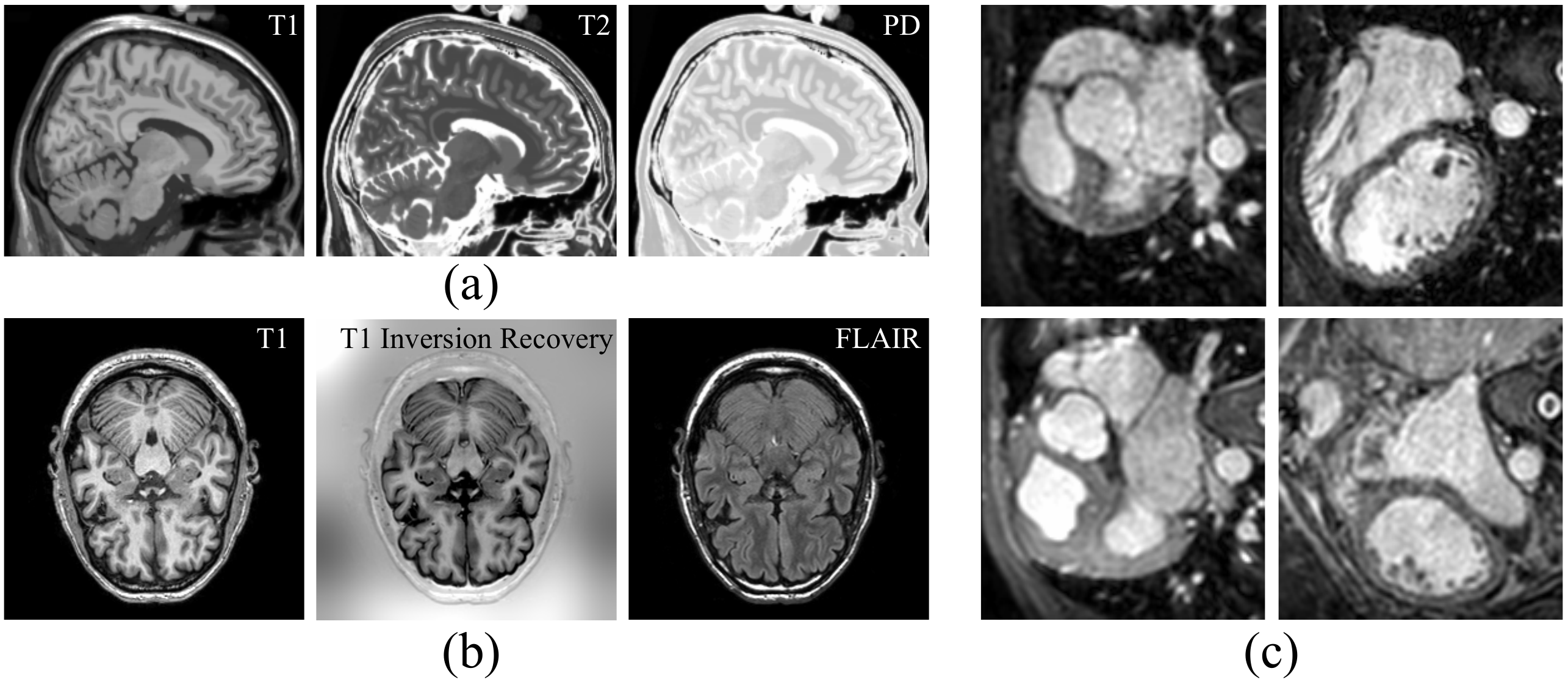}
	\caption{(a)-(c) Sample images from BrainWeb, MRBrainS and HVSMR, respectively.}
	\label{fig:datasets}
\end{figure}

	\paragraph{BrainWeb Dataset.} BrainWeb is a simulated database which contains one MRI volume for normal brain with three modalities: T1, T2 and PD. It contains 399 slices, among which we choose 239 slices for training and validation, and 160 for testing. The aim is to segment three tissues: cerebrospinal fluid (CSF), gray matter (GM), and white matter (WM). Images have the size of $217\times181$, $181\times181$ and $181\times217$ in three orthogonal views (see Figure \ref{fig:datasets}(a)). Skull stripping is conducted as the pre-processing technique before network training.
	
	\paragraph{MRBrainS Dataset.} MRBrainS contains T1, T1 inversion recovery and FLAIR sequences of real MR brain scans, among which 104 slices and 70 slices are utilized for training and testing from transversal view in our experiment. Each image is of size $240\times240$ with pixel-wise annotation (see Figure \ref{fig:datasets}(b)). Skull stripping is also conducted before network training.
	
	\paragraph{HVSMR Dataset.} HVSMR aims to segment blood pool and myocardium in cardiovascular MR images. We choose 10 MRI volumes and their ground truth annotation for network training and evaluation, among which 1868 slices and 1473 slices are utilized for training and testing (see Figure \ref{fig:datasets}(c)). There is no pre-processing before network training. 

	\paragraph{Implementation Details.} We concatenate the multiple modalities of MR slices as the input of our network for BrainWeb and MRBrainS. For HVSMR, since only one modality is provided, thus we utilize the single channel gray scale image as the input. As for evaluation metrics, we adopt Dice coefficient and Pixel Accuracy (PA) to quantitatively evaluate the segmentation performance. We use PyTorch as the implementation framework. An NVIDIA GeForce RTX 2080 is used for both training and testing. For training settings, we adopt batch normalization \cite{ioffe2015batch} after each convolutional layer. All the CNN backbones used in the following experiments share the same network structures proposed in \cite{simonyan2014very,he2016deep,ronneberger2015u}. We adopt a weight decay of $10^{-4}$ and use Adam \cite{kingma2014adam} for optimization, the learning rate starts from $6\times{10^{-4}}$ and gradually decays when training our CRDN.

\begin{table}[h!]
	\footnotesize
	\begin{center}
		\caption{Ablation study on BrainWeb, MRBrainS and HVSMR, evaluated by dice coefficient.}
		
		\label{tab:ablation}
		\begin{tabular}{p{0.16\columnwidth}c|c|c|c} % <-- Alignments: 1st column left, 2nd middle and 3rd right, with vertical lines in between
			\hline
			Decoder & Backbone & BrainWeb & MRBrainS & HVSMR\\
			\hline
			\multirow{3}{0.15\columnwidth}{RDC-ConvRNN} & VGG16 & 0.9927 & 0.9088 & \textbf{0.8813}\\
			& ResNet50 & 0.9920 & 0.9050 & 0.8641\\
			& U-Net-like & \textbf{0.9934} & 0.9068 & 0.8800\\
			\hline
			\multirow{3}{0.15\columnwidth}{RDC-ConvLSTM} & VGG16 & 0.9916 & \textbf{0.9126} & 0.8641\\
			& ResNet50 & 0.9896 & 0.9012 & 0.8606\\
			& U-Net-like & 0.9919 & 0.9112 & 0.8777\\
			\hline
			\multirow{3}{0.15\columnwidth}{RDC-ConvGRU} & VGG16 & 0.9926 & 0.9061 & 0.8776\\
			& ResNet50 & 0.9912 & 0.9021 & 0.8696\\
			& U-Net-like & 0.9925 & 0.9028 & 0.8796\\
			\hline
		\end{tabular}
	\end{center}
\end{table}

\begin{table*}[h!]
	\begin{center}
		\caption{Comparisons on BrainWeb, MRBrainS and HVSMR.}
		
		\label{tab:comparison}
		\begin{tabular}{c|c|c|c|c|c|c||c}
			\hline
			\multirow{2}{*}{Model} & \multicolumn{2}{c|}{BrainWeb} & \multicolumn{2}{c|}{MRBrainS} & \multicolumn{2}{c||}{HVSMR} & \# Params\\
			\cline{2-7}
			\multicolumn{1}{c|}{} & Pixel Acc & Dice & Pixel Acc & Dice & Pixel Acc & Dice & (240$\times$240$\times$3)\\
			\hline
			FCN with VGG16 & 0.9575 & 0.9142 & 0.9570 & 0.8637 & 0.9165 & 0.8368 & 50.42M\\
			SegNet with VGG16 & 0.9834 & 0.9679 & 0.9484 & 0.8294 & 0.8928 & 0.7718 & 29.45M\\
			U-Net with VGG16	& 0.9962 & 0.9923 & 0.9696 & 0.8991 & 0.9109 & 0.8201 & 25.86M\\
			\textbf{CRDN with VGG16}	& \textbf{0.9964} & \textbf{0.9927} & \textbf{0.9736} & \textbf{0.9126} & \textbf{0.9413} & \textbf{0.8813} & \textbf{14.87M}\\
			\hline
			FCN with ResNet50 & 0.9554 & 0.9115 & 0.9488 & 0.8374 & 0.9095 & 0.8266 & 115.83M\\
			U-Net with ResNet50	& 0.9954 & 0.9909 & 0.9710 & 0.9039 & 0.9192 & 0.8371 & 71.86M\\
			\textbf{CRDN with ResNet50}	& \textbf{0.9960} & \textbf{0.9920} & \textbf{0.9713} & \textbf{0.9050} & \textbf{0.9344} & \textbf{0.8696} & \textbf{23.65M}\\
			\hline
			FCN with U-Net backbone & 0.9579 & 0.9176 & 0.9564 & 0.8618 & 0.9179 & 0.8295 & \textbf{1.19M}\\
			SegNet with U-Net backbone & 0.9715 & 0.9455 & 0.9506 & 0.8448 & 0.9027 & 0.8099 & 2.36M\\
			U-Net & 0.9945 & 0.9892 & 0.9705 & 0.9021 & 0.9279 & 0.8593 & 1.94M\\
			\textbf{CRDN with U-Net-backbone}	& \textbf{0.9967} & \textbf{0.9934} & \textbf{0.9732} & \textbf{0.9112} & \textbf{0.9388} & \textbf{0.8800} & 1.23M\\
			\hline
		\end{tabular}
	\end{center}
\end{table*}

	\subsection{Ablation Study}
		In order to validate the effectiveness of RDC, as well as the performance of different combinations of CNN backbone encoders and RDC-based decoders for medical image segmentation, we first conduct an ablation study on above mentioned three datasets. We choose VGG16, ResNet50 and U-Net-like backbones as the feature encoder and use our three types of RDCs in the decoding phase. Note that VGG16 and ResNet50 are the ones reported in \cite{simonyan2014very,he2016deep} with batch normalization, and the U-Net-like backbone is the one reported in \cite{ronneberger2015u}, in which we compress the model scale by utilizing the channel number $\{16, 32, 64, 128, 256\}$ for each layer. All these model combinations are trained from scratch and tested on the whole dataset.

		The results of dice coefficient on three datasets are shown in Table \ref{tab:ablation}.  For brain datasets, we can see that RDC-ConvRNN with U-Net-like backbone performs the best on BrainWeb, while RDC-ConvLSTM with VGG16 backbone performs the best on MRBrainS. For HVSMR dataset, the RDC-ConvRNN with VGG16 backbone obtains the best performance, achieving $88.13\%$ dice coefficient value. The three types of RDCs all achieve relative high performance on brain datasets. Since BrainWeb is a simulated dataset and the intensity non-uniformity level is much lower than real brains, thus the relatively simple RDC-ConvRNN based decoder performs the best, but for more challenging dataset like MRBrainS, the RDC-ConvLSTM based decoder achieves much better segmentation results. For different use of CNN backbones, the results are similar, VGG16 and U-Net-like backbone achieve slightly better performance than ResNet50 in most cases and converge faster with residual block in our implementation. The results from three datasets indicate that encoders with deeper backbones are not often necessary for the task of medical image segmentation, yet our RDC-based decoder helps for hierarchical feature fusion.

\begin{figure}[htb]
	\centering
	\includegraphics[width=1.0\linewidth]{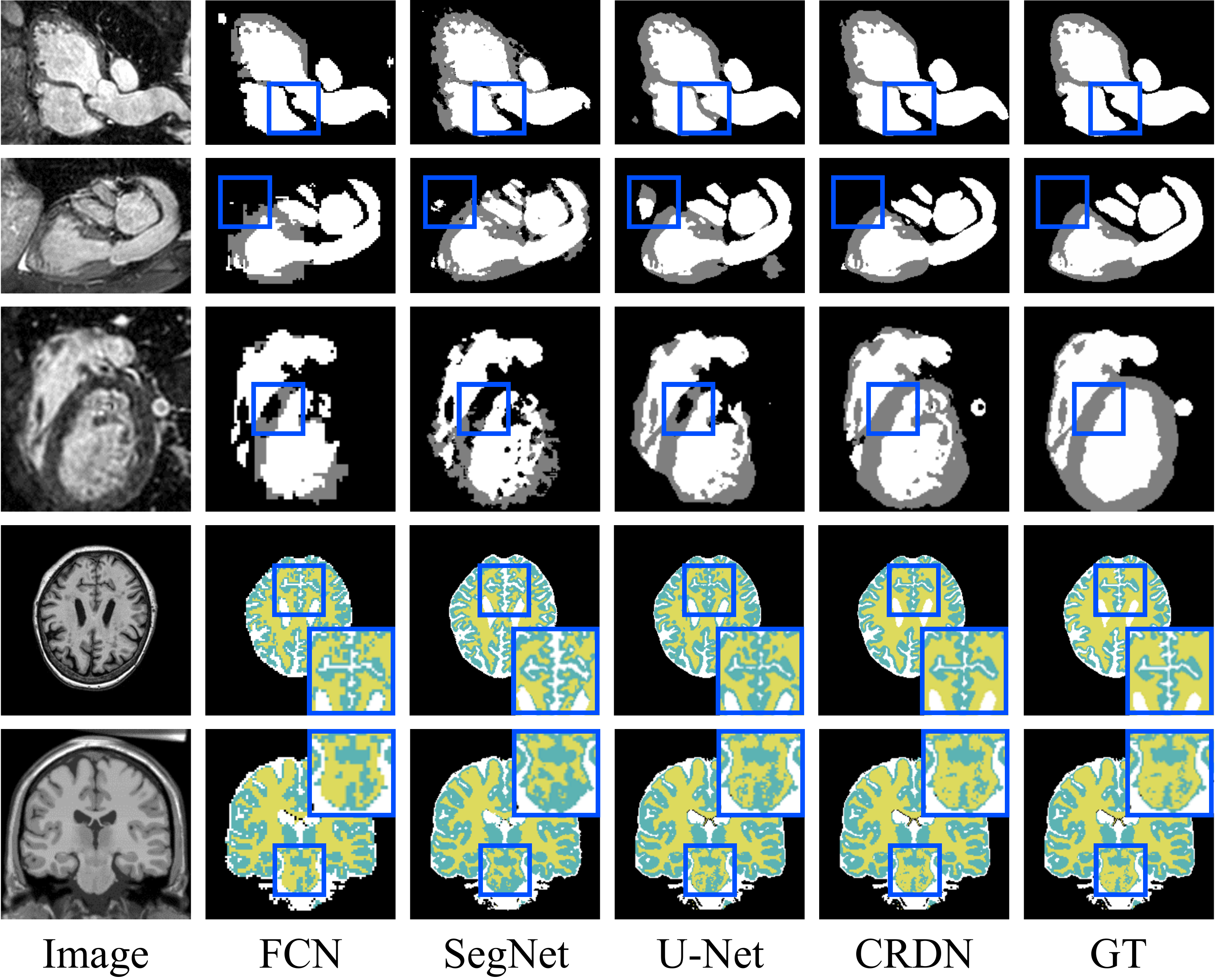}
	\caption{Some visualization results of the proposed CRDN and other encoding-decoding methods, i.e., FCN, SegNet and U-Net. All these methods utilize the U-Net-like backbone with different decoders. The top three rows are samples from HVSMR for segmenting two tissues (Blood Pool in gray, Myocardium in white), the fourth and the last rows are samples from MRBrainS and BrainWeb, respectively, for segmenting three tissues (WM in yellow, GM in green, and CSF in white). The blue rectangles highlight the noteworthy areas for comparisons.}
	\label{fig:vis}
\end{figure}

\begin{figure*}[htb]
	\centering
	\includegraphics[width=1.0\linewidth]{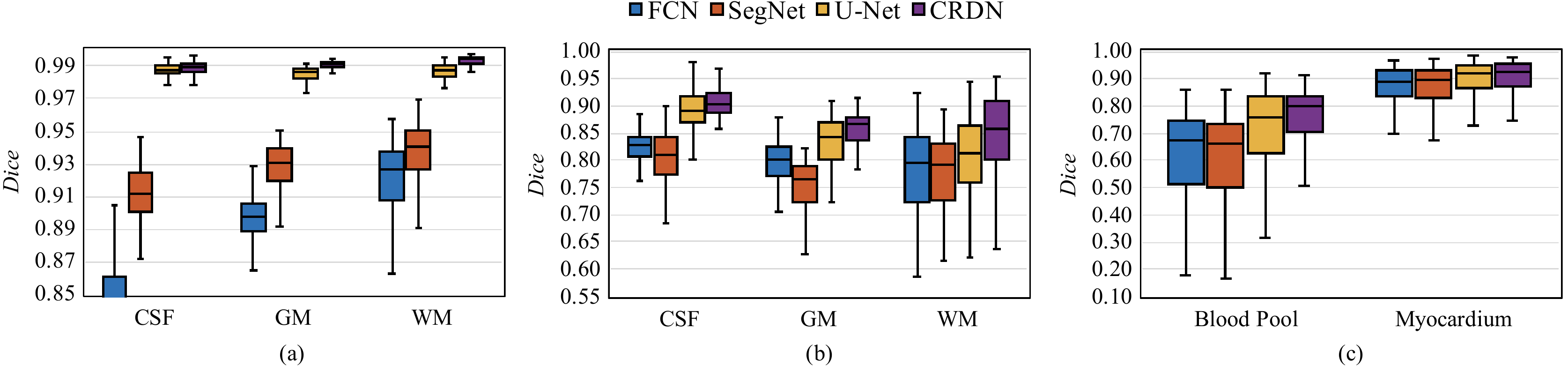}
	\caption{Boxplots of Dice coefficient for segmented tissues on BrainWeb, MRBrainS and HVSMR. Note that the three boxplots use different scales in the Y axes.}
	\label{fig:boxplot}
\end{figure*}

	\subsection{Comparison with Encoder-Decoder Networks}

		We further evaluate our method compared with the leading encoder-decoder models for medical image segmentation. We implement FCN, SegNet, U-Net and our CRDN with VGG16, ResNet50 and U-Net-like backbones as feature encoders. The FCN used in the experiment combines score maps from every layer of different spatial resolutions, and can also denote as ‘FCN-2s’, which obtains finer details than FCN-8s. We pick the best result of three RDC types to represent our CRDN model. The VGG16, ResNet50 and U-Net-like backbones contain 5-stage feature maps and are gradually combined with the 5 decoders. Note that SegNet with ResNet50 backbone is not implemented in our experiment because maxpooling is not used in each layer when downsampling the feature map.

		Table \ref{tab:comparison} shows the segmentation results on three datasets. We can find that CRDN achieves competitive results compared with other methods on all datasets. It is obvious that as the image data goes more complicated, the more superiority our model shows. The last column reveals the model size of different methods, and in the experiment, we choose a $240\times240\times3$ image as the input and compute the number of parameters used through each model. Since the parameters are shared through our CRDN, the trained models are much smaller while obtaining better segmentation performance compared with other encoder-decoder models. CRDN with U-Net backbone achieves competitive improvements on dice coefficient compared with U-Net --- $0.4\%$ relative improvements on BrainWeb, and $0.91\%$ on MRBrainS, $2.07\%$ on HVSMR. The number of parameters is only 1.23M which takes only about 0.01s for a single image with size $240\times240\times3$ in the testing phase. Figure \ref{fig:vis} illustrates some visualization results of the four models with U-Net-like backbone on three datasets. We can see that CRDN obtains finer details thanks to the memory mechanism in RNN for sequence processing. 

		Moreover, we analyze the segmentation performance over each tissue. Figure \ref{fig:boxplot}(a)(b) shows the dice value in terms of CSF, GM and WM tissues on BrainWeb and MRBrainS. The proposed method scores the highest median results for all of the tissues. The GM and WM obtain higher dice values than CSF on BrainWeb while the dice value of CSF improves a lot on MRBrainS. It also indicates that our method is superior to the other methods for segmenting WM, which accounts for a large proportion in human brains and, accordingly, achieves on average better segmentation results. Figure \ref{fig:boxplot}(c) shows the dice value in terms of blood pool and myocardium on HVSMR, the performance of the four methods for the myocardium segmentation is comparably the same — around $90\%$, yet our CRDN achieves better median results for segmenting blood pool.

\begin{figure}[htb]
	\centering
	\includegraphics[width=1\linewidth]{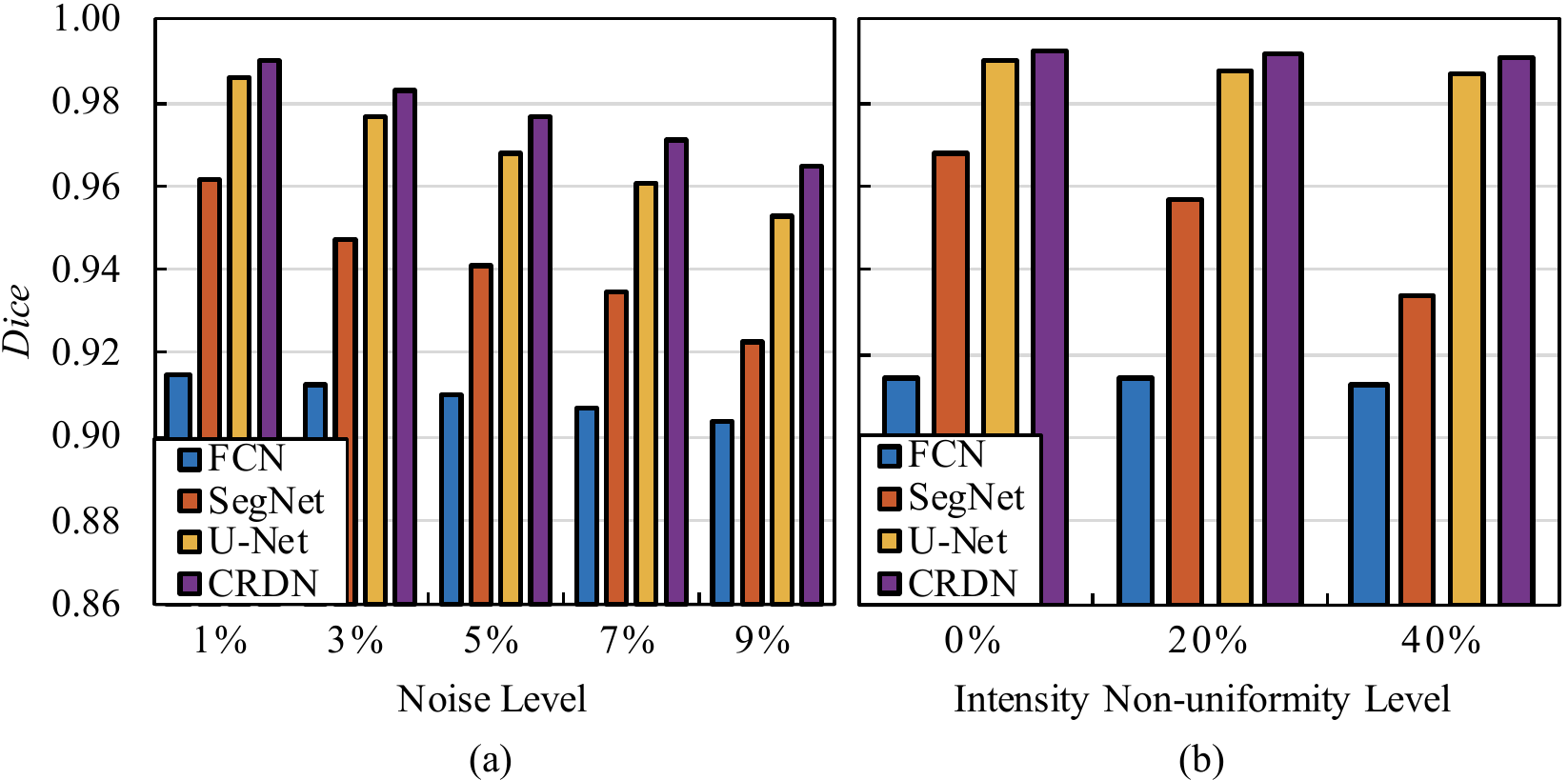}
	\caption{(a) Experimental results on images corrupted by noise. Note that the noise percentage represents the percent ratio of the standard deviation of the white Gaussian noise versus the signal for a reference tissue. (b) Experimental results on images affected intensity non-uniformity. Note that for a $20\%$ level, the multiplicative INU field has a range of values of $0.90, ..., 1.10$ over the brain area. For other INU levels, the field is linearly scaled accordingly \cite{cocosco1997brainweb}.}
	\label{fig:bar}
\end{figure}

	\subsection{Experiments on Network Robustness }

		Medical images, especially for MRI scans, are prone to image intensity-related artifacts such as noise and intensity non-uniformity (INU), which may be difficult for accurate visual inspection. Images are commonly affected by the white Gaussian noise due to the influence of magnetic field strength, and INU is mostly caused by RF excitation field inhomogeneity \cite{sled1998understanding}. To verify the model robustness to noise and INU, we compare CRDN with FCN, SegNet and U-Net on BrainWeb. BrainWeb provides MRI scans in 6 levels of noise and 3 levels of INU. All models are implemented based on the U-Net-like backbone training from scratch.

		\paragraph{Images Corrupted with Noise}  Figure \ref{fig:bar}(a) illustrates the decay of the segmentation results along with the increase of noise level. Note that the noise percentage represents the percent ratio of the standard deviation of the white Gaussian noise versus the signal for a reference tissue. From $1\%$ to $9\%$ of noise level, the dice coefficient reduces by $1.11\%$, $3.82\%$, $2.85\%$, and $1.69\%$ for FCN, SegNet, U-Net, and CRDN, respectively. Although FCN possesses the lowest decay of the four methods, the segmentation accuracy of FCN is far lower compared with other methods. Our CRDN does not drop much when facing strong noise and still performs the best among all other encoder-decoder networks.

		\paragraph{Images with Intensity Non-uniformity} Figure \ref{fig:bar}(b) illustrates the decay of the segmentation results along with the increase of INU level. The dice value of SegNet drops a lot when INU level becomes higher. The results of the four methods reduced by $0.13\%$, $3.41\%$, $0.28\%$, and $0.23\%$ from $0\%$ to $40\%$ of INU level. Our CRDN still keeps the dice coefficient over $99\%$ and is scarcely affected by non-uniformity intensities, which suggests that the proposed CRDN owns its robustness to intensity inhomogeneity for medical scans.

\section{Conclusion}
\label{sec:conclusion}
	In this paper, we propose the Recurrent Decoding Cell (RDC) for hierarchical feature fusion in encoder-decoder segmentation networks. The RDC combines the current score map of low resolution with the squeezed feature map of high resolution by leveraging the long-term memory capacity of convolutional RNNs. We also propose a Convolutional Recurrent Decoding Network (CDRN) based on RDC for multi-modality medical MRI segmentation. It utilizes a CNN backbone for feature extraction and the extracted feature maps from different layers are fed into a chain of RDC units to gradually recover the segmentation score maps. The experimental results demonstrate that RDC helps to achieve better boundary adherence compared with other segmentation decoders and reduces the model size. CRDN achieves promising segmentation results for medical image segmentation and shows its robustness to image noise and intensity non-uniformity in MRI. 

\section{Acknowledgements}
\label{sec:acjnowledgements}
	This work was supported in part by the National Nature Science Foundation of China (61773166, 61772369, 61672240), in part by the Natural Science Foundation of Shanghai (17ZR1408200), Joint Funds of the National Science Foundation of China (U18092006), Shanghai Municipal Science and Technology Committee of Shanghai Outstanding Academic Leaders Plan (19XD1434000), 2030 National Key AI Program of China (2018AAA0100500) and Projects of International Cooperation of Shanghai Municipal Science and Technology Committee (19490712800).

% References and End of Paper
% These lines must be placed at the end of your paper
\bibliography{2690_References}
\bibliographystyle{aaai}

\end{document}